\documentclass[letterpaper]{article}

\usepackage[T1]{fontenc}

\usepackage{geometry}
\usepackage{setspace}

\usepackage[style = chem-acs]{biblatex}
\usepackage{graphicx}
\usepackage{float}
\newfloat{scheme}{htbp}{los}
\floatname{scheme}{Scheme}
\floatname{chart}{Chart}
\newfloat{graph}{htbp}{loh}

\usepackage{chemformula} 
\usepackage[version = 4]{mhchem} 
\usepackage{upgreek}

\usepackage{xr}
\usepackage{amsmath, amssymb}
\usepackage{authblk}
\author[1*]{Matthew R. Wilson}
\author[1]{Benoit Guilhabert}
\author[1]{Jack A. Smith}
\author[1]{Michael J. Strain}
\author[1*]{Xavier Porte}
\affil[1]{Institute of Photonics, University of Strathclyde, Glasgow, G1 1RD, Scotland, UK}

\title{Chip-scale nanostructured chaotic billiards for broadband speckle spectrometry}

\date{*Email: matthew.wilson@strath.ac.uk, javier.porte-parera@strath.ac.uk}

\begin{document}

\maketitle

\begin{abstract}

Computational on-chip spectrometers are emerging as a powerful platform for portable spectral analysis, combining photonic integration with advanced signal processing to enable a wide range of in-situ sensing applications.
We propose a broadband reconstructive spectrometer based on wave chaos in a stadium microresonator with a nanostructured scattering layer for full-area speckle readout.
Wavelength-dependent interference within the chaotic microresonator encodes the spectral information into a spatial intensity pattern that can be computationally inverted to reconstruct the input spectra. 
The optimal fabrication parameters of the SU-8 polymer nanostructured layer yield a surface roughness of 176 nm and a root mean square thickness of $\simeq 2$ $\upmu$m. 
We experimentally validate our spectrometer at visible and infrared wavelengths, with resolutions of 43pm at 630nm and 8.2pm at 1550nm. 
The spectral reconstruction is demonstrated for single and multiple narrowline sources as well as for a broadband ($\sim 1$nm) pulsed laser source. 
The broad experimental validation and compact footprint (0.05mm$^2$) establishes our chaotic microresonator-based speckle spectrometer as a robust and versatile platform for high-resolution, on-chip spectral sensing.

\end{abstract}



\section*{Keywords}

On-chip spectrometer, chaotic billiards, speckle, nanograss, reconstructive spectrometer.

\section{Introduction}


The analysis of optical spectra is an essential task across many fields of science and technology \cite{Crocombe2018, Bec2021, Li2022}. 
This demand drives a growing interest in developing ever smaller and more cost effective spectrometers. 
Photonic integrated circuits (PICs) are a natural platform for realizing such miniaturized spectrometers owing to their compact footprint and scalable fabrication. 
In recent years there has been a surge of activity in developing on-chip computational spectrometers within the PIC community \cite{Peters2025}. 
Among the different miniaturization paradigms \cite{Yang2021}, reconstructive spectrometers (RS) have emerged as a versatile strategy for ready integrated systems. 
By introducing advanced computational analysis, RS can overcome the trade-off between footprint and performance that is typically encountered in other PIC-based spectrometers. 
RS rely on optical hardware to transform incoming light into a complex measurable signal, while leveraging advanced computer algorithms to decode that signal and reconstruct the original spectrum with high accuracy\cite{Xue2024, Zhang2025_Review}. 
Since its first demonstration in optical fibers \cite{Redding2012}, a variety of integrated photonic platforms have been proposed for reconstructive spectrometry \cite{Redding2013, Bao2015, Yang2019, Redding2016, Wang2019, Zhang:2025_speckle} with unprecedented levels of scalability and resolution. 

RS thrive in randomness, requiring a complex input to output relationship. 
The classic approach to generate such complex transfer matrix is by speckle patterns \cite{Cao2017}. 
Recently, wave chaos originating from a photonic integrated microresonator was proposed \cite{Zhang2025} as an alternative powerful mechanism for obtaining complex spectral transformations. 
This new mechanism led to the demonstration of a high-resolution spectrometer with an ultra-compact footprint. 
Chaotic microresonators are 2D waveguiding structures with geometries derived from the mathematical theory of chaotic billiards. 
The two main characteristics of chaotic microresonators are the high sensitivity to changes of the propagating light field (wavelength, phase and polarization) and their support of non-integrable ray trajectories \cite{Cao2015}. 
In the first demonstration of chaos-assisted spectrometry\cite{Zhang2025}, the internal field distribution of the resonator was coupled to a waveguide that acted as single output channel. 
Although the spectrometer was operated in time-multiplexed mode, a richer feature space could be obtained from accessing the full resonator's optical field. 

Here, we propose combining the high spectral sensitivity of chaotic microresonators with an area-scalable speckle readout mechanism based on nanoscale surface scattering. 
We deliberately introduce a disordered layer (known as 'nanograss' \cite{Adams2024}) via surface etching of a SU-8 polymer chaotic microresonator, enabling full-field optical sampling across the microresonator surface. 
This technique was recently proposed in polymer (SU-8) multimode waveguides for on-chip spectrometry \cite{Nafiz2024a, Nafiz2024b}. 
The nanograss layer efficiently scatters light from a waveguiding structure into free space, producing a speckle pattern that directly encodes spectral information into a spatial intensity distribution. 
These first demonstrations achieved high resolution and broadband operation, but required large propagation distances due to the straight waveguide geometry that limited the chip footprint. 
By capturing the spatially distributed response in our chaotic microresonators, the here proposed architecture accesses a substantially larger encoding space than conventional discrete readout schemes while retaining the long optical path lengths characteristic of chaotic geometries. 

In this work, we combine for the first time wave chaos and speckle spectrometry one a novel solution providing high spectral resolution (8.2 pm), large number of output channels (3025) and a very compact chip footprint (0.05mm$^2$). 
We first present a systematic optimization of SU-8 nanograss fabrication combining atomic force microscopy and optical characterization. 
This allowed us to maximize the scattered intensity at IR wavelengths while retaining the high degree of spectral sensitivity desired for our chaotic microresonators. 
Further, we characterized the spatial and spectral correlation dependencies of our speckle patterns. 
We finally describe the computational method used to recover input spectra from our experimentally measured speckle patterns and demonstrate their use for spectrometry. 
One crucial advantage of using SU-8 as a material platform is its broadband transparency covering visible and IR wavelengths and its relatively simple fabrication procedure.
Here, we demonstrate the ability of our proposed spectrometer to measure the spectra of single and multiple narrowline sources at IR (1550nm) and visible (630nm) wavelengths and a broadband source in the IR, obtaining optical resolutions of 43pm at 630nm and 8.2pm at 1550nm. 
The spectral reconstruction is demonstrated for single and multiple narrow linewidth sources as well as for a broadband ($\sim 1$nm) pulsed laser source. 
Speckle spectrometers based on nanograssed chaotic microresonators offer an advanced yet robust monolithic platform for broadband, high-resolution on-chip spectroscopy with ultra-compact footprint.


\section{Results and Discussion}

We fabricated a set of chaotic microresonators using direct-write laser lithography. 
The fabrication process is detailed in the Methods section. 
We used the canonical example of a chaotic geometry, which is the Bunimovich stadium \cite{Bunimovich1979, Cao2015}. 
Recently, stadium microresonators have attracted the interest of the neuromorphic computing community due to the highly complex nonlinear dynamics associated with wave chaos and their extreme spectral sensitivity \cite{UchidaSunadaSciRep2019, Wilson2025, You2025}. 

\begin{figure}[H]
    \centering
    \includegraphics[width=1\linewidth]{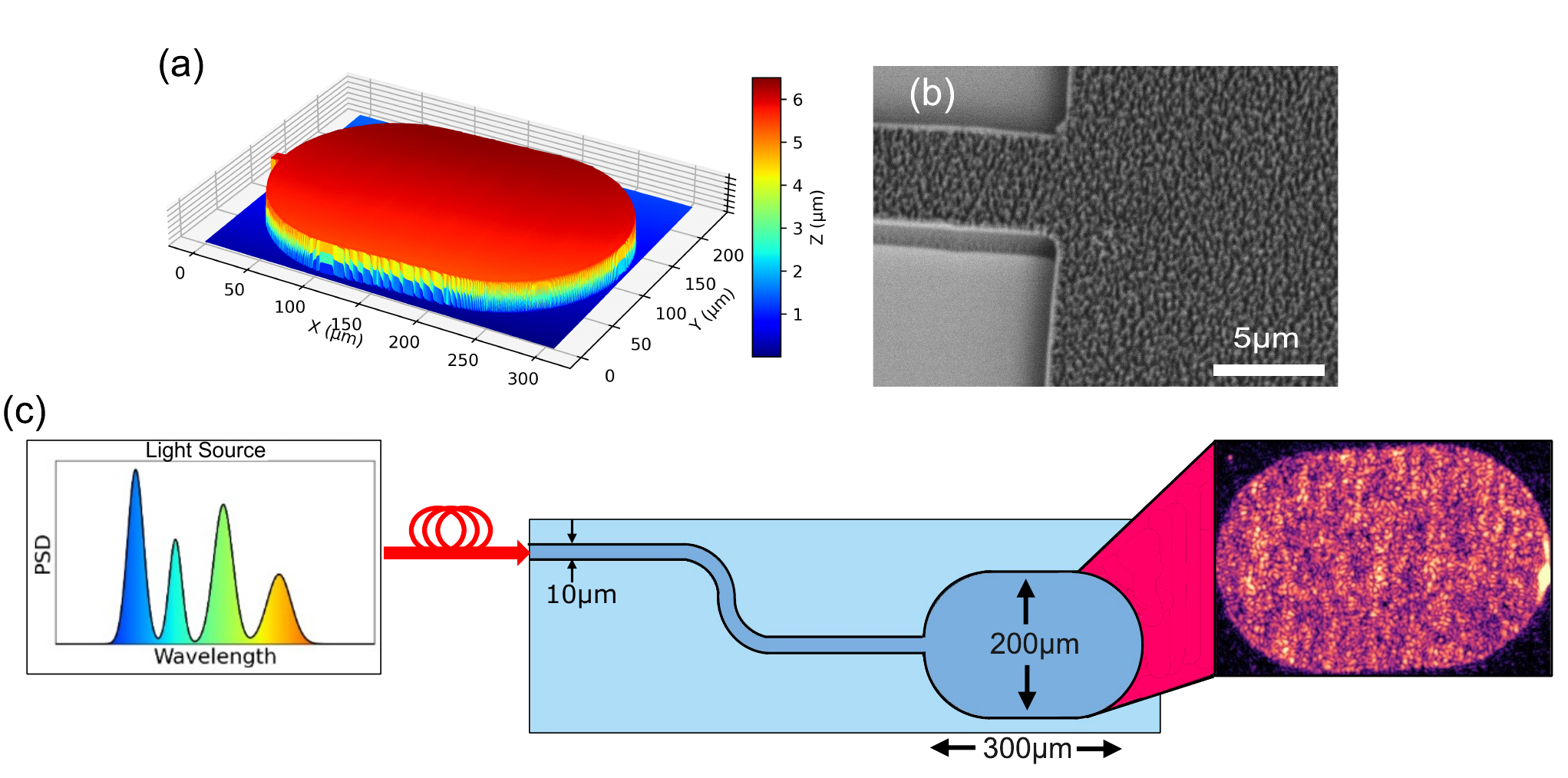}
    \caption{Qualitative imaging examples and spectrometer schematic. 
    (a) Optical profilometer scan of an unetched stadium. 
    Colour map denotes the height in $\upmu \mathrm{m}$ units. 
    (b) SEM of a stadium after a 6 minute etch process, showing clear nanograss formation. 
    (c) Schematic of the spectral to spatial transformation performed by the on-chip chaotic microresonator. 
    The insert image on the right is a measured InGaAs camera image of the device at a single wavelength injection at 1550nm.}
    \label{fig:intro}
\end{figure}

The Bunimovich stadium is formed by two semicircles separated by a rectangular section and are defined by the ratio of the semicircle radius R to the straight edge length of the rectangular section L. 
Each stadium was designed to have a radius-to-length ratio of 1:1, meaning that the surface area is $A=(\pi+2)R^2$. 
Fixing R = 100$\upmu m$ gives a surface area of 0.05mm$^2$, which is kept consistent across the fabricated devices. 
Each stadium is fed by a multimode waveguide with a width of $10\upmu \mathrm{m}$, located centrally on the left side of the microresonators. 
Figure \ref{fig:intro}a shows an example stadium microresonator after hard baking, imaged using an optical profilometer (Veeco Wyko NT1100).  
In typical on-chip optical components, light scattering from the top surface is minimal. 
In order to access the full cavity's optical field, we use an O$_2$ plasma reactive ion etch (cf. Methods section for detailed description) to roughen the top surface. 
The scanning electron microscope (SEM) image in Fig. \ref{fig:intro}b illustrates the randomly distributed scattering structure that results of this process, which is known as nanograss \cite{Adams2024, Nafiz2024a, Nafiz2024b}. 
The nanograss on the surface of the device accentuates scattered light, transforming spectral information into a spatially distributed speckle pattern as shown in Fig. \ref{fig:intro}c. 

\subsection{Characterization of the surface nanostructuring process}

\begin{figure}[H]
    \centering
    \includegraphics[width=0.95\linewidth]{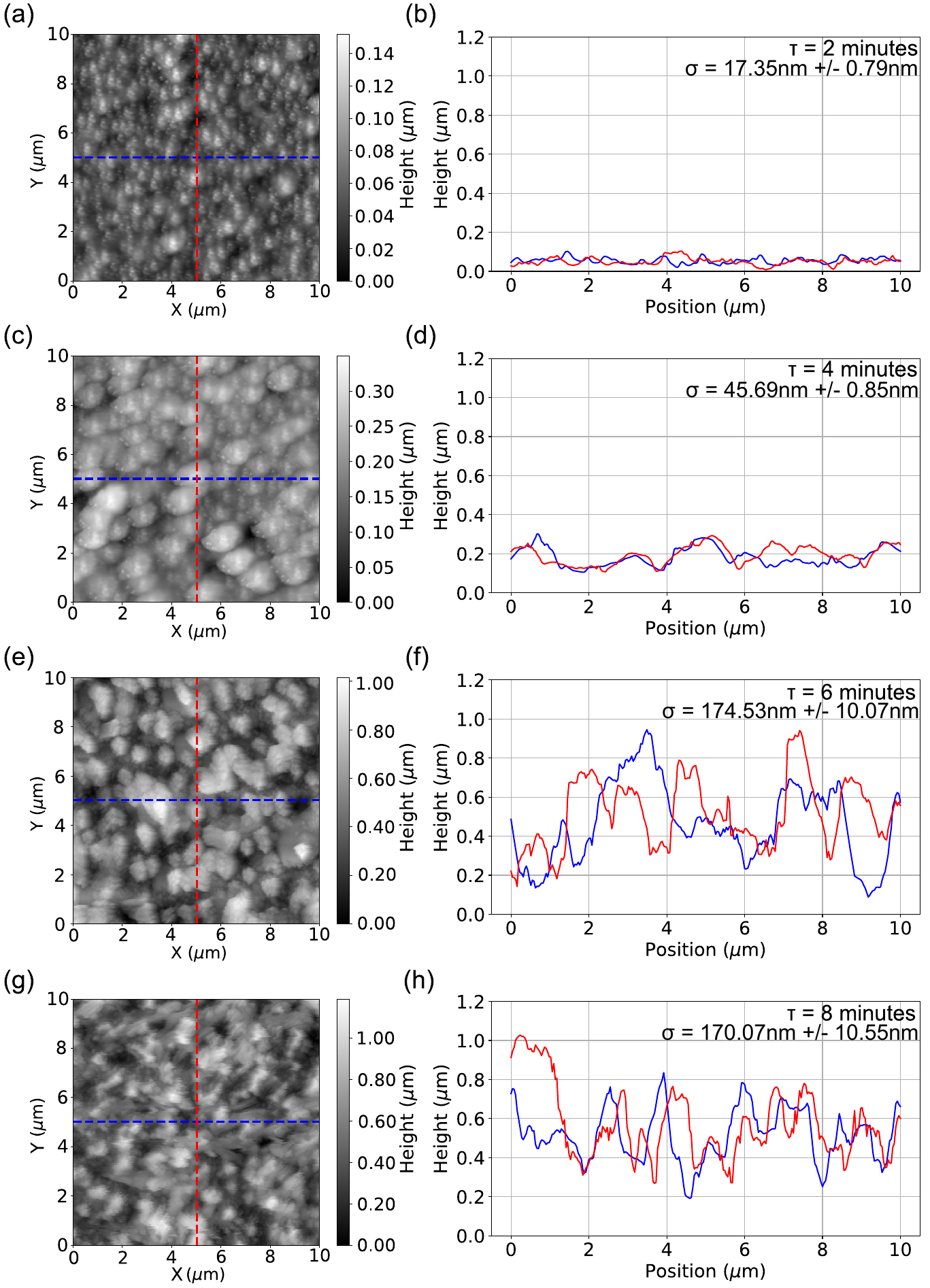}
    \caption{Examples of topology scans over a $10 \upmu \mathrm{m} \times 10\upmu \mathrm{m}$ area located at the center of each stadium device, for chips with etch times 2 minutes (a,b), 4 minutes (c,d), 6 minutes (e,f), and 8 minutes (g,h). 
    Left side panels depict the surface map of the height in grayscale. 
    The vertical (red) and horizontal (blue) line cuts are plotted in the right side panels.}
    \label{fig:AFM_2}
\end{figure}

The nanograss fabrication process was first characterized using non-contact atomic force microscopy (NC-AFM). 
High resolution topology scans were performed over a $10 \upmu \mathrm{m} \times 10 \upmu \mathrm{m}$ area near the center of each microresonator in order to characterize the material's surface as a function of etch time $\tau$, i.e. the time for which the SU-8 is exposed to the plasma. 
Representative scanning results are shown in Fig. \ref{fig:AFM_2} for etching times of 2 minutes (a,b), 4 minutes (c,d), 6 minutes (e,f), and 8 minutes (g,h). 
The left panels in Fig. \ref{fig:AFM_2} illustrate the morphological evolution of the SU-8 material as the etching progresses in time. 
The horizontal (blue) and vertical (red) cuts in each image are respectively plotted in the right side panels. 
The change in morphology is accompanied by a sharp increase of the root mean square (RMS) surface roughness $\sigma$, which is here the average of 3 different devices with identical etch times per chip. 
After 2 minutes of etching, the surface still remains relatively smooth, with surface roughness of 17.35nm $\pm$ 0.79nm, enabling only limited optical scattering. 
The scenario significantly differs for etching times beyond 4 minutes. 
For etching times of 6 and 8 minutes, the surface topology presents in-plane features at length scales comparable to the operating wavelength (peaking at 174.53nm $\pm$ 10.07nm) enhancing the nanograss' impact on the optical scattering.  

\begin{figure}[H]
    \centering
    \includegraphics[width=1\linewidth]{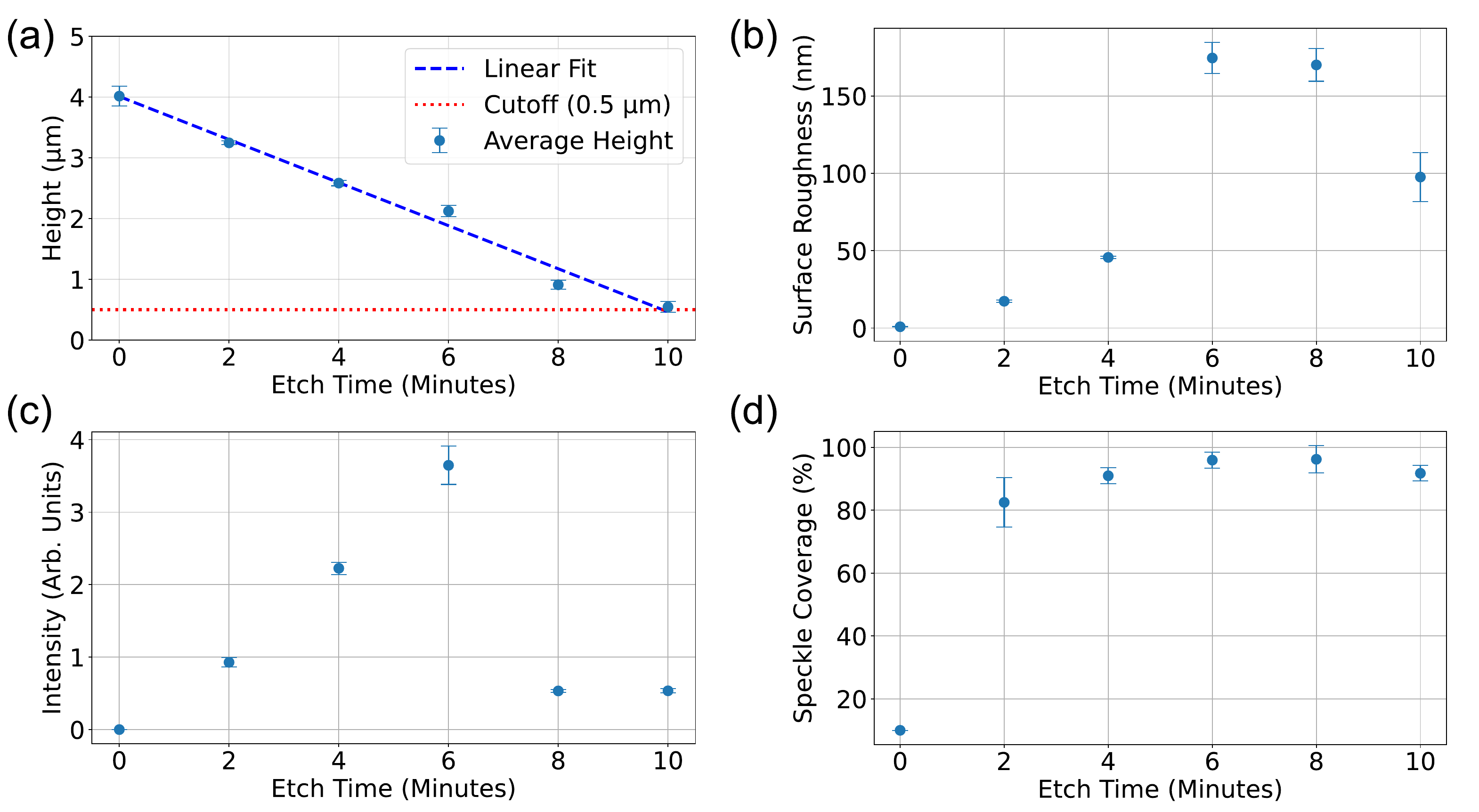}
    \caption{Characterization results. 
    (a) Height of SU-8 devices recorded by NC-AFM measurements. 
    Data shows average height measured across multiple devices per etch time with standard deviation error bars. 
    Linear fit to the data returns an etch rate of 355nm $\pm$ 22nm per minute. 
    The red dotted line denotes the $\simeq 0.5 \upmu \mathrm{m}$ cut-off height for guided modes at 1550nm. 
    (b) RMS surface roughness measured by NC-AFM, with standard deviation error bars. 
    (c) Average speckle intensity measured across the identical cavities with standard deviation error bars as a function of etch time. 
    (d) Speckle coverage measurement. 
    Fraction of the stadium microresonators containing a measurable speckle pattern as a function of etch time.}
    \label{fig:characterization}
\end{figure}

Plasma etching dramatically modifies the surface, affecting the optical properties of our photonic microcavities. 
In order to optimize the optical properties, we measured the effects of etch time on four quantifiers: Etching depth, surface roughness, scattered light intensity, and spatial coverage of the speckle patterns.
Figure \ref{fig:characterization}a shows the average height measured for different devices versus etch time, with error bars that correspond to the standard deviation of the height between devices with the same etch time. 
We determine the height of the waveguiding layer by measuring the step height at the microresonator edge using NC-AFM. 
This allowed us to determine that the average height of the unetched devices after UV lithography and hard baking was 4.02$\upmu$m $\pm$ 0.16$\upmu$m. 
This drops to an average device height of $0.55 \upmu \mathrm{m} \pm 0.09 \upmu \mathrm{m}$ after 10 minutes of etching. 
Fitting a linear decay to the data, we were able to determine that the etch rate of our RIE recipe was approximately 355nm $\pm$ 22nm per minute. 
The height cut-off for guided modes at 1550nm in SU-8 is $\simeq 0.5 \upmu \mathrm{m}$ calculated using finite element method (FEM) mode simulation, denoted by the red dotted line in Fig. \ref{fig:characterization}a. 
The average RMS surface roughness across each chip is depicted in Fig. \ref{fig:characterization}b. 
Overall, we observe an exponential increase in roughness as we etch the waveguiding layer, up to a saturation maximum around 6 minutes etch time. 
Beyond this point we gradually smoothen the surface again, in agreement with recent results of etching SU-8 waveguides \cite{Adams2024}.

We complemented the NC-AFM characterization with a direct measurement of the light scattered from the microcavity using the optical setup introduced in supplementary Fig. S1. 
The light from a C-band tuneable laser source (TLS) is side-coupled to the chip via a single mode lensed fiber. 
Any light vertically scattered by the nanograss is then collected via imaging optics in the setup and measured by an InGaAs camera (Triton SWIR, Lucid Vision Labs) with a global shutter acquisition and an integration time of 13ms. 
The InGaAs camera has a detector area of 1280 pixels $\times$ 1024 pixels (1.3 megapixel) with a pixel pitch of 5$\upmu m$. 
The presence of surface roughness increases propagation loss in integrated optical devices, caused by the increased amount of light being scattered out from the guiding layer \cite{Lacey-Payne_loss_model:1994}. 
From this, we can assume that our highest detected intensity should be seen in the same cavities where the surface roughness peaks. 
This is confirmed in Fig. \ref{fig:characterization}c, where we observe a clear peak intensity measured in the IR camera corresponding to an etch time of 6 minutes, which in-turn corresponds to the etch time of the highest RMS surface roughness. 
We also note a sharp drop in the measured pixel intensity for devices etched beyond this time. 
This can be explained from the strong interaction of the single mode light propagation with the wavelength-magnitude surface roughness, as characterized in Figs. \ref{fig:characterization}a and \ref{fig:characterization}b. 
Any speckle pattern seen in those cavities will be dominated by stray light from the input fiber interacting with the remanent sub-guiding SU-8 layer.
The last magnitude we measured is the fraction of cavity surface covered by speckles of similar intensity, which we refer as speckle coverage. 
To quantify the distribution of intensities across the stadium surface we measure the percentage of the image that has pixel values above a background threshold. 
The background threshold is determined by measuring the average pixel intensities of dark regions in the images that are known to not overlap with the microresonator. 
Figure \ref{fig:characterization}d shows the result of this measurement normalized to the microresonator area. 
The speckle coverage peaks close to $\sim 100\%$ at 6 minutes etch time, confirming that this etch time presents the optimum optical behavior.  

\begin{figure}[H]
    \centering
    \includegraphics[width=1\linewidth]{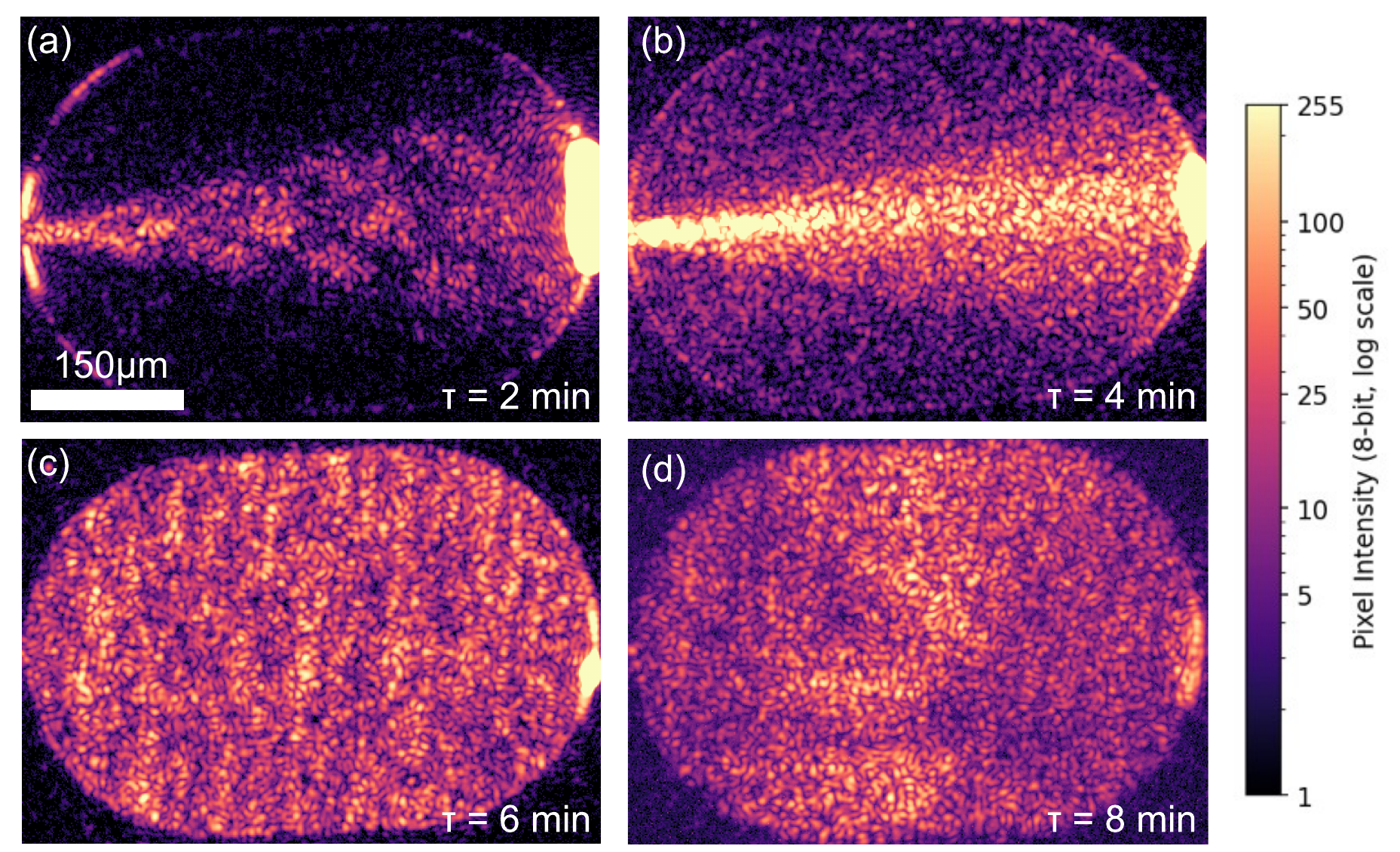}
    \caption{Exemplary grayscale images of speckle patterns observed via nanograss for stadiums etched (a) 2 minutes, (b) 4 minutes, (c) 6 minutes, and (d) 8 minutes. Images are shown in log scale with a false color map to aid visualization.}
    \label{fig:coverage}
\end{figure}

Figure \ref{fig:coverage} depicts the structural changes to the measured speckle patterns for devices with different etching times and provides qualitative illustration of the speckle coverage measured in Fig. \ref{fig:characterization}d. 
The grayscale images are plotted in log scale with a false color map to aid visualization. 
All images were acquired using the same camera settings, however as already discussed in Fig. \ref{fig:characterization}c, the measured intensity of the speckle patterns has a dependence on the etch time. 
For that reason, each of the images shown in \ref{fig:coverage} were acquired with different input powers to aid visualization. 
By exposing the devices to the O$_2$ plasma for only 2 minutes (Fig. \ref{fig:coverage}a), we are able to primarily detect the initial diffraction cone of the light entering the microresonator and the initial reflection point of the light at the stadium boundary. 
After 4 minutes of etching  (Fig. \ref{fig:coverage}b), the initial diffraction cone starts to degrade and speckles from previously dark regions become clearly visible. 
As we etch further the diffraction cone completely disappears from the speckle pattern and a more even distribution of intensities is detected as seen in Fig. \ref{fig:coverage}c, corresponding to an etch time of 6 minutes.

\subsection{Spatial and spectral speckle correlations}

We have shown that the spatial distribution of the speckle pattern across the microresonator surface is dependent on etch time via surface roughness and cavity height. 
To complete the picture on the functionality of our stadium microcavities as speckle spectrometers, we must study the spatial and spectral correlation properties of the resulting speckle patterns. 
For this, we first studied the underlying structure in the speckle patterns using 2D spatial autocorrelation. 
This is a standard measurement in speckle statistics and is outlined mathematically in Eq. \ref{eqn:autocorr}. 
The optical intensity at position $\textbf{r}(x,y)$, with $|\textbf{r}|=\sqrt{x^2+y^2}$, is denoted by I(\textbf{r}). 
The brackets $\langle \cdot \rangle$ represent the radial spatial average and $\sigma_I$ is the standard deviation of intensity values. 

\begin{equation}
C(\Delta \mathbf{r}) = \frac{\left\langle I(\mathbf{r}) \, I(\mathbf{r} + \Delta \mathbf{r}) \right\rangle}{\sigma_I^2}
\label{eqn:autocorr}
\end{equation}

\begin{figure}[H]
    \centering
    \includegraphics[width=1\linewidth]{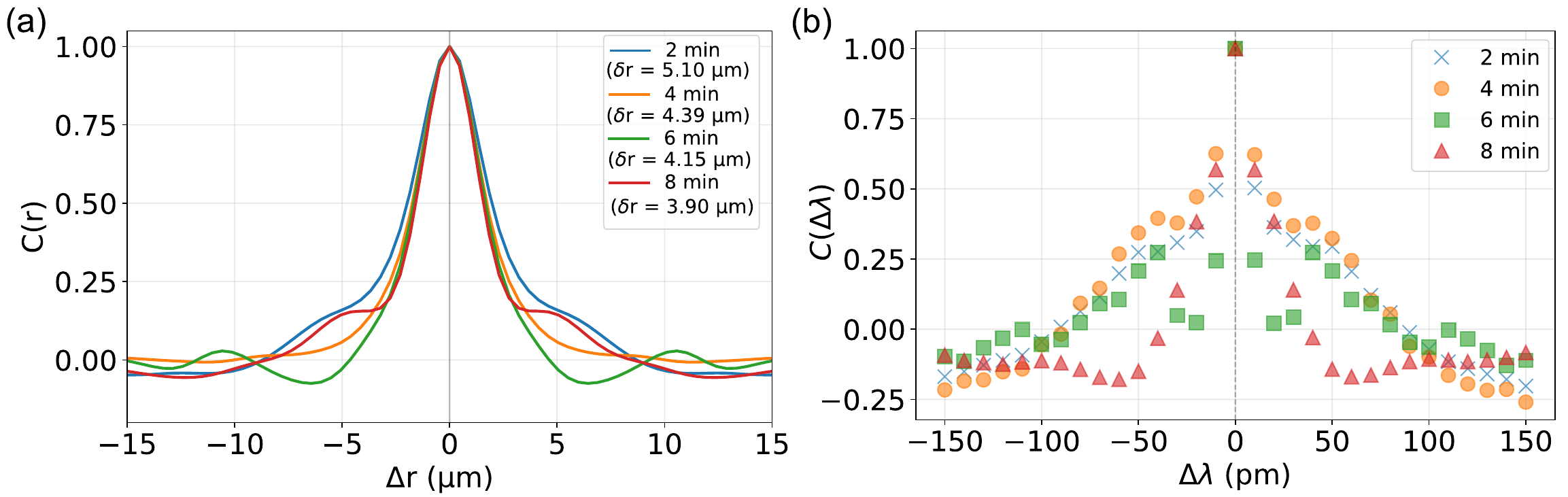}
    \caption{
    Spatial and spectral correlation dependencies of speckle patterns as a function of etch time. (a) Autocorrelation of speckle patterns from cavities etched 2, 4, 6 and 8 minutes with $\lambda=1550nm$. $\delta$r denotes the decay half-width at $C(\Delta r)=\frac{1}{e}$. (b) Spectral correlation function for the different etch times.}
    \label{fig:corr_etch}
\end{figure}

The autocorrelation characterizes the statistical correlation between intensity fluctuations at two points separated by a radial displacement $\Delta \textbf{r}$, and is normalized such that $C(0) = 1$. 
Figure \ref{fig:corr_etch}a shows the spatial autocorrelation for speckle patterns corresponding to cavities injected with a wavelength of 1550nm. 
The overall shape of those autocorrelation curves are non-trivial and reflect the complex interaction between post-etch surface topology and stadium cavity geometry. 
Although the detailed study and modeling of the correlation curves shapes is beyond the scope of the current work, we can straightforwardly extract the speckle correlation length $\delta \mathrm{r}$  as the half-width half maximum (HWHM). 
This magnitude provides us with a statistical measurement of how separated two independent speckle grains are. 
From the comparison of the different curves in Fig. \ref{fig:corr_etch}a, we observe that $\delta \mathrm{r}$ systematically decreases as we increase etching time. 
In practice, this indicates us how to appropriately subsample the speckle pattern spatially to maximize the information in the readout while keeping computation efficient. 
For example, this can inform the minimum pitch for the pixels of a detector array in an heterogeneously integrated system. 

To evaluate the viability of our nanostructured stadium cavities for computational spectrometry, we also measure the spectral decorrelation of the speckle patterns. the spectral correlation follows the following expression

\begin{equation}
C(\Delta\lambda) = \frac{\left\langle \big(I(\lambda)-\langle I(\lambda)\rangle\big) \, \big(I(\lambda+\Delta\lambda)-\langle I(\lambda+\Delta\lambda)\rangle\big) \right\rangle_r}{\sigma_I(\lambda) \, \sigma_I(\lambda+\Delta\lambda)},
\label{eqn:norm_cov}
\end{equation}

\noindent where $I(\lambda)$ is the speckle intensity distribution at wavelength $\lambda$, $\Delta\lambda$ is the wavelength separation between the two patterns, and $\sigma_I(\lambda)$ is the standard deviation of the speckle intensity. 
The brackets $\langle\cdot\rangle_r$ denote the spatial average over the detector pixels. 
By construction, the normalized covariance satisfies $C(0)=1$. 
Finally, we calculate the mean of the correlation coefficients corresponding to equal wavelength step sizes.

Figure \ref{fig:corr_etch}b shows the spectral correlation $C(\Delta\lambda)$ on speckle patterns acquired during a stepped wavelength sweep over a range of 4nm with a step size of 0.01nm, using Eq. \ref{eqn:norm_cov}. 
Before calculating the spectral correlation between speckle patterns we crop the image to a region in the center of the microresonators with dimensions $100 \text{ pixels} \times 100 \text{ pixels}$, corresponding to an area of $46\upmu m \times 46\upmu m$. 
Within this region, we perform an algorithmic masking of low variance pixels to remove saturated pixels, which withholds approximately 5\% of pixels from the correlation measurement per speckle image. 
From Fig. \ref{fig:corr_etch}b, we see that the width of the decay falls as the devices are etched, demonstrating that as we create a more even distribution of intensities in the speckle pattern, we are able to access more of the complex spectral behavior of the microresonator. 
Interestingly the spectral correlation decay begins to broaden again for the devices etched for 8 minutes. 
As previously noted, as the guiding layer thickness reduces, it increases the mode overlap with the scattering layer and that reduces the effective lifetime of the light in the microresonator. 
This is in good correspondence with this measured broadening of the spectral correlation. 
The HWHM of a spectral correlation decay curve provides a measurement of spectral resolution $\delta\lambda$ \cite{Cao2013,Peters2025}. 
To extract a value of $\delta\lambda$ with higher precision, we repeated the measurement of Fig. \ref{fig:corr_etch}b for two extra wavelength sweeps with finer step sizes of 5pm and 1pm. 
This process helped us to fill in the gap between the first and second points on the spectral correlation plot as well as to rule out artifacts from possible wavelength undersampling process. 
The results of this study are shown in supplementary Fig. S2. 
We report a final spectral resolution of 8.2pm, belonging to the device etched for 6 minutes. 
From this study we extract that the optimum etch time to maximize readout intensity and spectral variability is 6 minutes, corresponding to an RMS surface roughness of 176nm, a device height of $\sim 2 \upmu \mathrm{m}$ and $\delta\lambda$ of 8.2pm.

We additionally investigated the effects of area scaling on the spectral decorrelation of the speckles, 5 stadium microresonators were fabricated with different surface areas while maintaining the same ratio of length to width. 
For each device, the spectral correlation was measured as before and $\delta\lambda$ was extracted from the resulting curves (see supplementary Fig. S3). 
As already discussed, the spectral resolution of a speckle spectrometer is fundamentally linked to the rate at which the speckle pattern decorrelates with wavelength step size, $\delta\lambda$ provides a direct experimental measure of the achievable resolution. 
Across all fabricated devices, only small variations in $\delta\lambda$ were observed, with no clear systematic dependence on microresonator area. 
The spectral decorrelation width has a variance of only $0.1 \mathrm{pm}^2$, which can be approximated as remaining constant as the stadium dimensions are scaled. 
This indicates that simply increasing the physical area of the microresonators does not significantly alter the spectral sensitivity of the speckle patterns. 
This ability to scale the system is particularly advantageous for practical implementations of a chaotic microresonator speckle spectrometer. 
The microresonators area could be tailored to match detector pixel sizes or increase the total number of independent spatial channels available for reconstruction. 
Additionally, we prove that smaller devices keep equally competitive resolutions, which is very attractive for further fully integrated applications. 
Since the spectral correlation width remains consistent, these geometric modifications can be implemented without necessitating any major changes to the optics or the reconstruction methodology.

\subsection{Speckle Spectrometry}

The reconstruction algorithm can be formulated as a linear inverse problem, closely related to the training stage of photonic extreme learning machines (ELMs)\cite{Biasi2023, Wilson2025}, in which wavelength-dependent speckle patterns generated by the chaotic microresonator encode spectral information into a spatial intensity distribution. 
Due to the strong wavelength sensitivity of interference within the microresonators, each wavelength produces a distinct speckle pattern, enabling a high-dimensional mapping from spectrum to measured intensity. 
A calibration dataset is first acquired by recording speckle images at discrete, known wavelengths spanning the spectral range of interest using a narrow linewidth TLS. 
Each two-dimensional speckle image is spatially downsampled into square regions of interest (ROIs) of side length $\delta r$, chosen to match the measured speckle correlation length of $2\upmu \mathrm{m}$. 
The intensity within each ROI is summed, such that each resulting feature corresponds approximately to an independent speckle grain. 
The number of ROIs for a stadium microcavity is 3025, which is equivalent to the number of spectral readout channels. 
These features are then flattened into a one-dimensional vector. 
Stacking the vectors corresponding to all calibration wavelengths column-wise yields the calibration matrix $\mathbf{H} \in \mathbb{R}^{N \times m}$, where $N$ is the number of spatial features and $m$ is the number of calibrated spectral channels. 
Each column of $\mathbf{H}$ represents the system response to a single wavelength.
For an unknown input spectrum, a single speckle image is measured and processed identically to produce a feature vector $\vec{y}$. 
The forward model is then given by 

\begin{equation}
    \vec{y} = \mathbf{H}\vec{\beta}, 
\label{eqn:forward}
\end{equation}

where $\vec{\beta}$ represents the unknown spectral intensity distribution over the calibrated wavelength grid. 
The problem that we want to solve is therefore recovering $\vec{\beta}$ from $\vec{y}$. 
In order to do so, we compute a regularized inverse of $\mathbf{H}$ using a truncated singular value decomposition (SVD), as defined by 

\begin{equation}
    \mathbf{H} = \mathbf{U}\boldsymbol{\Sigma}\mathbf{V}^T
    \xrightarrow{}
    \mathbf{H}^{+} = \mathbf{V}_k \boldsymbol{\Sigma}_k^{-1} \mathbf{U}_k^T .
\label{eqn:svd}
\end{equation}

Here, \textbf{U} and \textbf{V} are orthogonal matrices whose columns contain the left and right singular vectors of \textbf{H}, respectively. 
The diagonal matrix $\boldsymbol{\Sigma}$ contains the singular values s$_i$ in descending order. 
The singular values quantify the strength of the independent measurement channels present in the calibration matrix. 
In the truncated SVD, only the  largest singular values and their corresponding singular vectors are retained, yielding the regularized pseudoinverse \textbf{H}$^{{+}}$. 
The truncation is defined by retaining singular values satisfying $s_i \geq c_t s_{\max}$, where $c_t$ is the truncation threshold, and $k$ denotes the number of retained singular values. 
This truncation suppresses noise amplification associated with small singular values and acts as an implicit regularization. 
An initial estimate of the spectrum is then obtained from 

\begin{equation}
    \vec{\beta}_{\mathrm{SVD}} = \mathbf{H}^{+} \vec{y}.
\label{eqn:beta}
\end{equation}

While this method is robust, it often results in noisy reconstructions \cite{Cao2013}. 
We therefore add a nonlinear refinement step in the form of a support-restricted non-negative least-squares (NNLS) optimization. 
The support set S is first defined by selecting the indices corresponding to the largest components of $\vec{\beta}_{\mathrm{SVD}}$. 
The refined spectrum is then obtained by solving 

\begin{equation}
\vec{\beta}_{S} = \arg\min_{\vec{\beta} \geq 0} \left\| \mathbf{H}_{S} \vec{\beta} - \vec{y} \right\|_2^2 ,
\label{eqn:nnls}
\end{equation}

where $\mathbf{H}_{S}$ contains only the columns of $\mathbf{H}$ indexed by S. 
The final reconstructed spectrum is obtained by embedding $\vec{\beta}_{S}$ back into the full wavelength grid. 
This support-restricted NNLS step reduces noise, enforces non-negativity, and suppresses artifacts while preserving the dominant spectral components identified by the SVD estimate.\\

For validation, the reconstructed spectrum is compared against the equivalent spectrum measured with an optical spectrum analyzer (OSA). 
The agreement between the two provides a qualitative measure of reconstruction fidelity. 
Physically, the chaotic microresonator acts as a wavelength-dependent random encoder, mapping spectral information into a high-dimensional speckle pattern. 
The reconstruction process corresponds to a regularized inversion of this mapping, where truncated SVD ensures stability and the NNLS refinement enforces physically meaningful solutions. 
This approach enables accurate single-line and multi-line spectral reconstruction without requiring tunable optical components, relying solely on passive interference within the microresonators. 
In simple terms, the calibration process creates a library of speckle patterns, where each pattern corresponds to a known wavelength. 
When an unknown spectrum is measured, the resulting speckle image can be viewed as a combination of the speckle patterns produced by all wavelengths present in the input. 
The reconstruction algorithm determines the set of wavelength contributions that, when combined, best reproduces the measured speckle image. 
The truncated SVD provides an initial estimate of these contributions while suppressing noise, and the NNLS refinement then adjusts the amplitudes to obtain a physically meaningful spectrum containing only non-negative intensities. 
The output spectrum therefore represents the wavelengths and relative powers that most likely generated the observed speckle pattern.

First, the ability of our system to determine which wavelengths are present in the signal from narrow linewidth laser sources was characterized. 
Light from a C-band TLS (TSL570, Santec) was injected into the chaotic microresonator and stepped in wavelength by 0.01nm over a 1nm wavelength range from 1548nm to 1549nm. 
The TLS has a set-point accuracy of $\pm5pm$. 
The window size of 1nm was selected for the demonstration due to experimental constraints regarding the stepping speed of the TLS. 
The calibration sweep was performed with a step size of 10pm, which defines our spectral channel spacing and is close to $\delta\lambda$. 
At each wavelength step, an image of the resulting speckle pattern was acquired and labeled according to the set-point of the TLS. 
These labeled speckle images formed the training dataset. 
The OSA used for verification of the reconstruction is a Yokogawa AQ6374 optical spectrum analyzer with a resolution bandwidth of 0.05nm. 
Our test data set was created by performing the same wavelength sweep, but this time combining the TLS with a constant wavelength DFB laser source. 
The DFB source has a measured wavelength of 1548.75nm $\pm$ 0.05nm. 

\begin{figure}[H]
    \centering
    \includegraphics[width=1\linewidth]{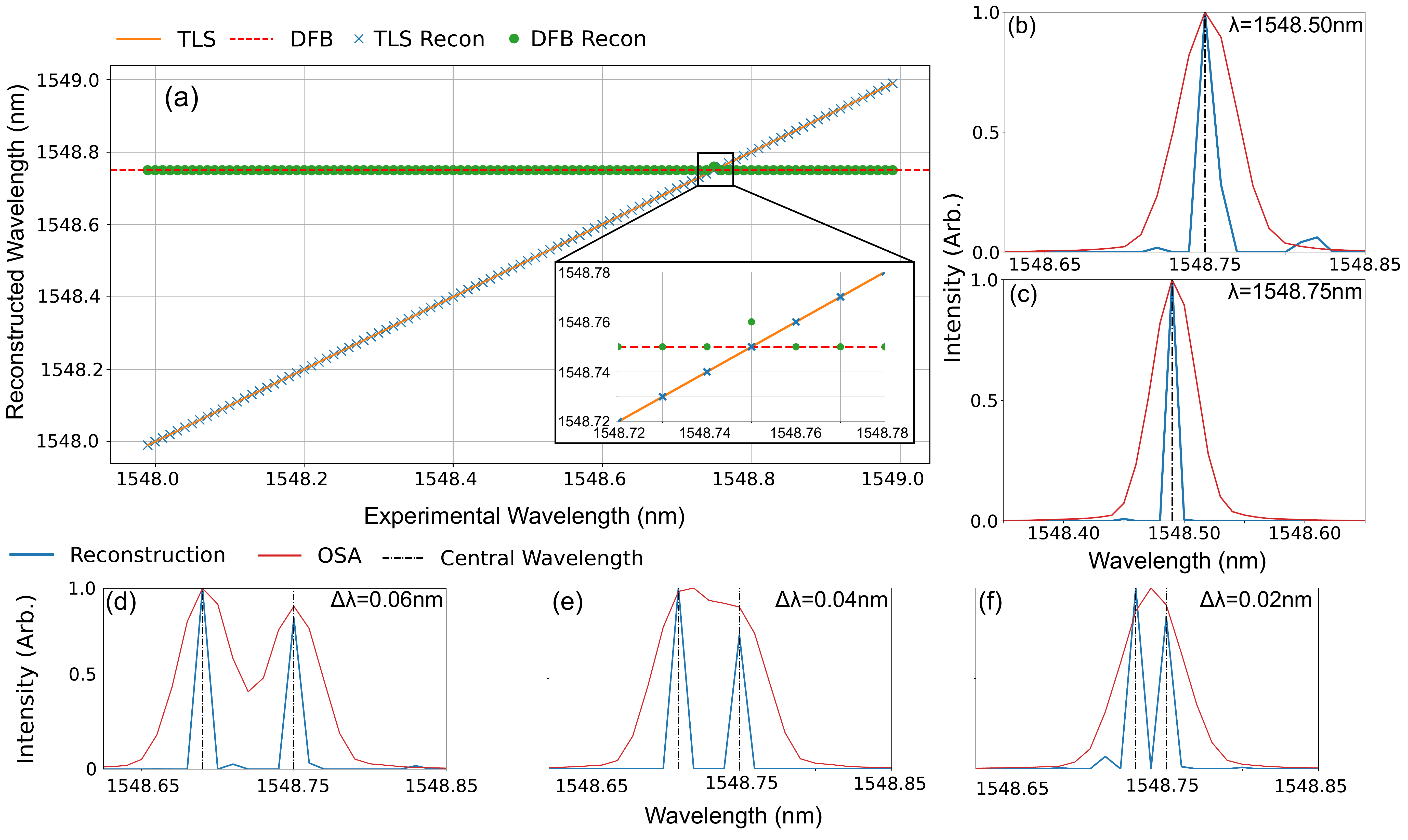}
    \caption{Quantitative and qualitative characterization of spectral reconstruction in the IR. 
    (a) Reconstruction accuracy over a 1nm range when the microresonator is injected by two laser sources simultaneously. 
    The solid orange line and the dashed red line denote the set-point wavelength of the TLS and the measured wavelength of the DFB respectively. 
    The blue crosses and green circles denote the dominant wavelengths present in the spectral reconstruction respectively. 
    (b) Example of single line reconstruction of the TLS source at a central wavelength of 1548.50nm. 
    (c) Example reconstruction of the DFB source. 
    (d-f) Reconstructions for the two laser sources with corresponding spectral separations of (d) 0.06nm, (e) 0.04nm and (f) 0.02nm.}
    \label{fig:spec_recon}
\end{figure}

Figure \ref{fig:spec_recon}a depicts the peak wavelengths of the two lasers while scanning the TLS. 
From each image, we extract the wavelength value of the two largest peaks from the reconstructed spectra and plot them against the expected wavelengths (the set-point wavelength of the TLS and the measured wavelength of the DFB using the OSA). 
This analysis shows that the algorithm returns the expected central wavelength for both the TLS and DFB sources with high fidelity. 
The only error occurs when the spectral separation of both lasers drops below the channel resolution limit $\delta\lambda$, as shown at the inset of Fig. \ref{fig:spec_recon}a. 
At this point the reconstruction does still identify a secondary peak, however it is located 10pm away from the experimental wavelength limited by the resolution of the method. 
Full spectral reconstructions are also presented in Fig. \ref{fig:spec_recon}b-f where the blue lines represent the reconstructions, the red lines represent the measured traces from the OSA and the black dashed lines represent the expected central wavelength. 
Figure \ref{fig:spec_recon}b and \ref{fig:spec_recon}c show the reconstructions based on speckle patterns produced by injecting the microresonators with the TLS and DFB respectively. 
Figure \ref{fig:spec_recon}d-f shows the reconstruction based on two-line injection presented in Fig. \ref{fig:spec_recon}a as the spectral line of the TLS approaches that of the DFB. As can be seen, we are able to resolve not only the correct central wavelength of the two lines, but their relative intensities as confirmed by the OSA verification measurement (red line). It should be noted that reconstructions are shown within a cropped wavelength window. This is to aid in visualizing the reconstruction, all other data points outwith this window are zero.

\begin{figure}[H]
    \centering
    \includegraphics[width=1\linewidth]{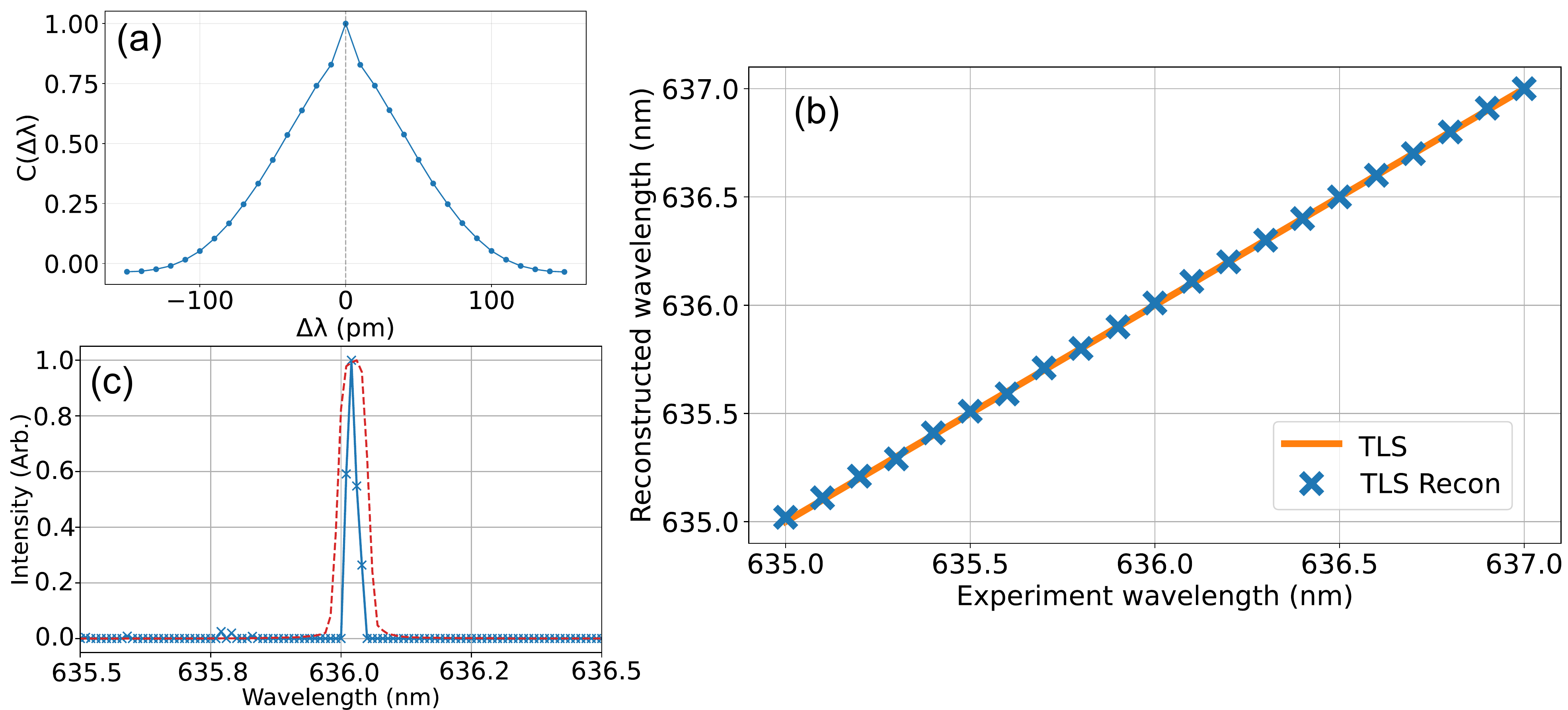}
    \caption{Spectrometer characterization in the visible domain. 
    (a) Correlation of speckle patterns as a function of wavelength step size for 635nm TLS. 
    (b) Wavelength extracted from reconstruction of a wavelength sweep within a 2nm band starting from 635nm. 
    The blue crosses represent the extracted peaks from the reconstruction algorithm and the orange solid line represents the expected wavelength based on the laser set-point. 
    The mean absolute wavelength error was calculated to be 4pm over this band. 
    The central wavelength recovered from every $10^{th}$ reconstruction steps is shown for clarity. (c) Reconstruction example.}
    \label{fig:red}
\end{figure}

A principal benefit of using SU-8 as a device material is its inherent broadband transparency \cite{Cassells24}. 
This means that a single device can interface with both established telecoms wavelengths and visible wavelengths. 
Here, we demonstrate the recovery of spectra within the band 635nm to 637nm using the same device used in the reconstruction of IR spectra. 
We begin by injecting the microresonators with light from a red TLS (Newport Velocity TLB-6704) which is continuously swept at a rate of 0.2nm per second. 
For measurements at red we substitute the InGaAs camera for a CMOS camera (U3-3682XLE-NIR, IDS) with a detector area of 2592 pixels $\times$ 1944 pixels (5.04 megapixel) and a pixel pitch of 2.2$\upmu m$. 
This camera has a rolling shutter and an integration time of 20ms was used. 
A video of the sweep was captured at a rate of 20 FPS, leading to a wavelength sampling interval of 0.01nm. 
Before checking the spectral reconstruction ability of the system, we first check the spectral correlation decay to determine the theoretical resolution of the spectrometer for red light. 
The results of this are presented in Fig. \ref{fig:red}a, showing a correlation decay with $\delta\lambda$ of 43pm. Having established the theoretical spectral resolution of the computational spectrometer, we tested the wavelength accuracy for narrowline reconstruction. As shown previously for the IR spectral range, we extract the wavelength of the dominant peak in the reconstruction and plot it against the expected wavelength as shown in Fig. \ref{fig:red}b. 
The mean absolute wavelength error $\varepsilon_\lambda = \overline{||\lambda_{recon}-\lambda_{true}||}$ over the 2nm band was calculated as 4pm. 
An example of narrowline reconstruction is shown in Fig. \ref{fig:red}c. 
These results demonstrate that the algorithm is able to use the speckle pattern at red wavelengths to recover spectra with a greater resolution than our commercial bench-top OSA. 
As with the IR reconstructions, Fig. \ref{fig:red}c is shown within a cropped spectral window. 
Again, all data points outside this window are zero.\\

\begin{figure}[H]
    \centering
    \includegraphics[width=1\linewidth]{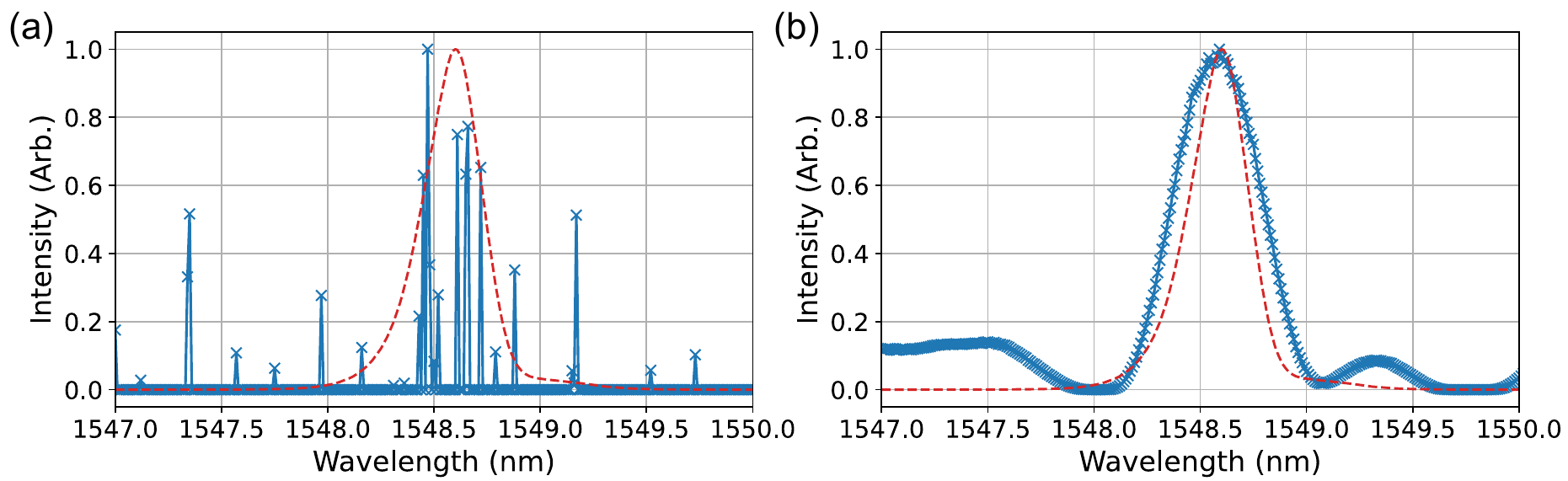}
    \caption{IR broadband source reconstruction. 
    (a) Failed reconstruction of a broadband source resulting from the NNLS method used in narrowband reconstruction. 
    (b) Example successful reconstruction of a broadband source using ridge regression method. 
    Blue crosses denote the reconstruction and red dashed line shows experimentally measured OSA trace.}
    \label{fig:broadband}
\end{figure}

We finally evaluate the performance of the spectrometer when exposed to a broadband optical source in the IR spectral region. 
The microresonator was injected using a pulsed laser source operating in CW mode (UOC-3, Pritel Inc) with light centered at a wavelength of 1548.60nm with an approximate spectral width of $\sim 1 \mathrm{nm}$. 
We observe that spectral reconstruction generally fails when the least-squares approach, outlined in Eq. \ref{eqn:nnls}, is applied to broadband reconstruction. 
While the method of least-squares reconstruction is robust for narrowband spectra, its performance is poor for broadband inputs due to the reduction in speckle contrast and increased correlation between speckle patterns associated with neighboring wavelengths. 
An example of a failed reconstruction using this method is presented in Fig. \ref{fig:broadband}a. 
Similar limitations of least-squares–based reconstruction for broadband signals have been reported in speckle-based spectrometers \cite{Zhang2025, Zhang:2025_speckle}. 
To address this limitation, we replace the NNLS optimizer with a modified ridge-regression that incorporates a one-dimensional biharmonic (graph-Laplacian-squared) regularization on the output weights, which has been shown to stabilize broadband spectral reconstruction by enforcing smoothness along the wavelength axis \cite{Zhang:2025_speckle,Zhao:2025_diffreg}. 
The resulting optimization equation can be written as 

\begin{equation}
\vec{\beta}_{\mathrm{ridge}} = \arg\min_{\vec{\beta}\geq 0} \left(\left|H_S\vec{\beta}-\vec{y} \right|_2^2 + \beta \left| \mathbf{D}_2 \vec{\beta} \right|_2^2 \right),
\label{eqn:ridge}
\end{equation}

where $\mathbf{D}_2 \in \mathbb{R}^{(N-2)\times N}$ is the second-order finite-difference operator acting on the wavelength index. 
Its action on the output-weight vector is defined component-wise as

\begin{equation}
\left(\mathbf{D}_2 \vec{\beta}\right)_i = \beta_i - 2\beta_{i+1} + \beta_{i+2},
\qquad i = 1,\dots,N-2.
\label{eqn:D2_def}
\end{equation}

The resulting spectral reconstruction is shown in Fig. \ref{fig:broadband}b. 
Using this method we achieve a reconstruction in good agreement with our verification OSA measurement, with some evidence of overestimation in the spectral wings.

Notably, a limitation of the reported experiment is that the absolute wavelength accuracy is fundamentally limited by the accuracy of the initial calibration. 
Any systematic offset with respect to the calibration of the TLS or drifts in the environmental conditions will be transferred directly to the reconstructed spectra, although relative spectral features remain consistent. 
For this, the stability of the proposed spectrometer was additionally studied to support the presented results and is reported in the supplementary material. 
To quantify environmental drifts in our experiment, we performed back-to-back stepped sweeps of the IR and red TLSs and measured the cross correlation between speckle patterns corresponding to like wavelengths.
The time interval between the start of each sweep was 8 minutes in the IR and 100 seconds in the red as this is the respective time taken by the TLSs to execute a sweep  over the test ranges.
The results of this are depicted in supplementary Fig. S5 and Fig. S6. 
This study showed a temporal instability that we attribute to thermal drift in the laboratory temperature.
In order to characterize in detail the temperature variability of the speckle patterns, we heat the chip substrate with a temperature controlled stage (Peltier plus thermistor). 
The microresonators were injected at a constant wavelength of 1550nm and an image was acquired for each step in temperature after allowing the speckle pattern to stabilize. 
The correlation was then measured and the resulting plot is shown in supplementary Fig. S4. 
The HWHM of this decays shows that the speckle patterns are stable to within $\sim$0.5$^o$C.
As a result, in order to keep a consistent performance with respect to environmental changes, the temperature of the microresonator must be kept constant within $\sim$0.5$^o$C range after initial calibration.

\section{Conclusions}

We have demonstrated a broadband computational spectrometer based on a chaotic stadium microresonators fabricated in SU-8 polymer with a nanograss scattering layer to enable scalable readout of internal interference patterns. 
A combination of NC-AFM and optical characterization identified an optimum etch condition corresponding to an RMS surface roughness of 176 nm and a device thickness of $\simeq 2 \upmu \mathrm{m}$, providing optimal detection of up-scattered light while preserving the high degree of wavelength sensitivity and speckle coverage associated with chaotic microresonators. 
We measured spectral resolutions based on correlation curves of 43pm at 630nm and 8.2pm at 1550nm. 
We demonstrated computational reconstruction of single and multiple narrow linewidth spectral sources at both visible and IR wavelengths using the same device. 
We additionally showed the spectral reconstruction of a broadband optical source with a linewidth of $\sim 1 \mathrm{nm}$. 
Together, these results establish polymer nanostructured chaotic billiard microresonators as a highly compact ($\simeq 0.05 \mathrm{mm}^2$) and versatile platform for computational spectrometry. 

\section{Methods}

\subsection{Fabrication}

Glass microscope slides are solvent cleaned in acetone, methanol and IPA for 3 minutes each. 
They are then rinsed in DI water and allowed to dehydrate on a hotplate with a temperature $>110\,^\circ\mathrm{C}$ for 10 minutes. 
The glass substrate is then spin coated with a layer of SU-8 6005 using a spin profile of 500 RPM for 5 seconds then ramping to 3000 RPM for 60 seconds, targeting a post-exposure thickness of 5$\upmu$m. 
The sample is then soft-baked for 3 minutes at $110\,^\circ\mathrm{C}$. 
The device designs are then written into the soft-baked SU-8 using a UV direct write laser lithography tool (Heidelberg DWL66+). 
The exposed SU-8 is then given a post-exposure bake at $110\,^\circ\mathrm{C}$ for 3 minutes before developing in undiluted PGMEA. 
The SU-8 is placed in PGMEA for 1 minute and quenched in IPA for 10 seconds before being returned to the PGMEA for a further 20 seconds and IPA for 20 seconds in 10 second intervals to fully remove the unexposed resist. 
After an inspection of the device quality in an optical microscope, the developed SU-8 devices are hard-baked at $180\,^\circ\mathrm{C}$ for 20 minutes. 
The SU-8 nanograss layer is achieved by applying an O\textsubscript{2} plasma in a reactive ion etcher (Oxford Plasma Instruments) with an accelerating power of 200 W and a flow rate of 30 sccm at a pressure of 0.1 Torr.

\section{Supporting Information}

Supporting information: Additional experimental details and methods, including schematic of the experimental setup (PDF)

\section*{Acknowledgements}

The authors thank Dr Paul Edwards for his assistance in SEM imaging. 
The authors thank Prof Martin Dawson for helpful discussions. 

\section*{Funding} 

The Volkswagen Foundation; Engineering and Physical Sciences Research Council; Fraunhofer UK. 

\section*{Disclosures} 

The authors declare no conflicts of interest.



\printbibliography

\newpage

\section*{Supporting Information for "Chip-scale nanostructured chaotic billiards for broadband speckle spectrometry"}

\setcounter{figure}{0}
\renewcommand{\thefigure}{S\arabic{figure}}

\subsection*{S1: Experimental Setup}

In this section, we explain the optical imaging setup used in this work. 
The example below is shown for IR experiments, but the laser source and camera are interchangeable with the equivalent visible components. 
Light is coupled from a infrared (IR) C-band tunable laser source (TSL570, Santec) via a lensed fiber onto our chip. 
A multimode waveguide carries the in-coupled light into the stadium microresonator. 
The spatially distributed intensities are then captured via the imaging optics (Nikon N10X-PF, AC254-150-C-ML) and measured with an InGaAs camera (Triton SWIR, Lucid Vision Labs). 
When stepping wavelength (e.g. for calibration of the spectrometer) the camera provides an electrical trigger to the TLS to step wavelength after an image is acquired. 
The recorded images are saved as bitmap files in the laboratory PC. 
Note that the LED in the imaging column is for broadband illumination of the chip. 
The LED is only used for alignment purposes, and during measurements it is switched off. 

\begin{figure}[H]
    \centering
    \includegraphics[width=0.75\linewidth]{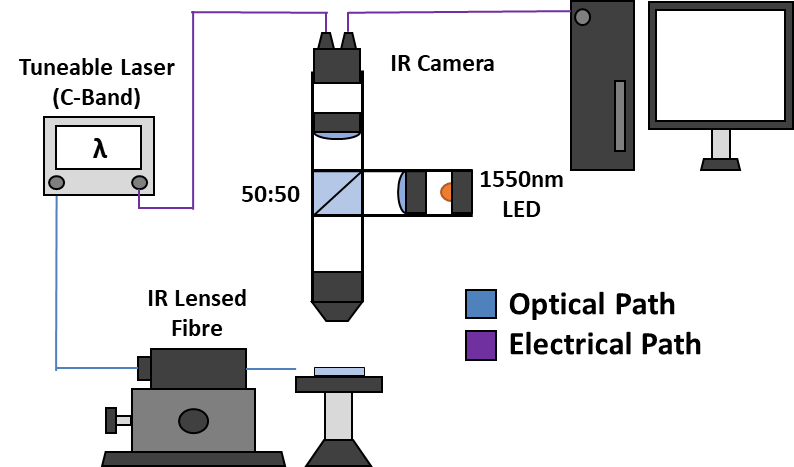}
    \caption{Schematic of the experimental setup}
    \label{fig:setup}
\end{figure}

\newpage

\subsection*{S2: Highly-resolved spectral correlation}

We calculated the spectral correlation curves for different resolution steps of the c-band tunable laser source. 
We perform this measurement only for the optimal etch time of 6 minutes, due to experimental constraints related to the stepping speed of the tunable laser source.
Here, we compare sweeps with spectral resolutions of 10pm (green triangles), 5pm (orange squares) and 1pm (blue circles). 
As can be seen, using these shorter wavelength step sizes fills in the curve presented in the main text (the green triangles), not affecting to the final value of $\delta\lambda$. 

\begin{figure}[H]
    \centering
    \includegraphics[width=0.6\linewidth]{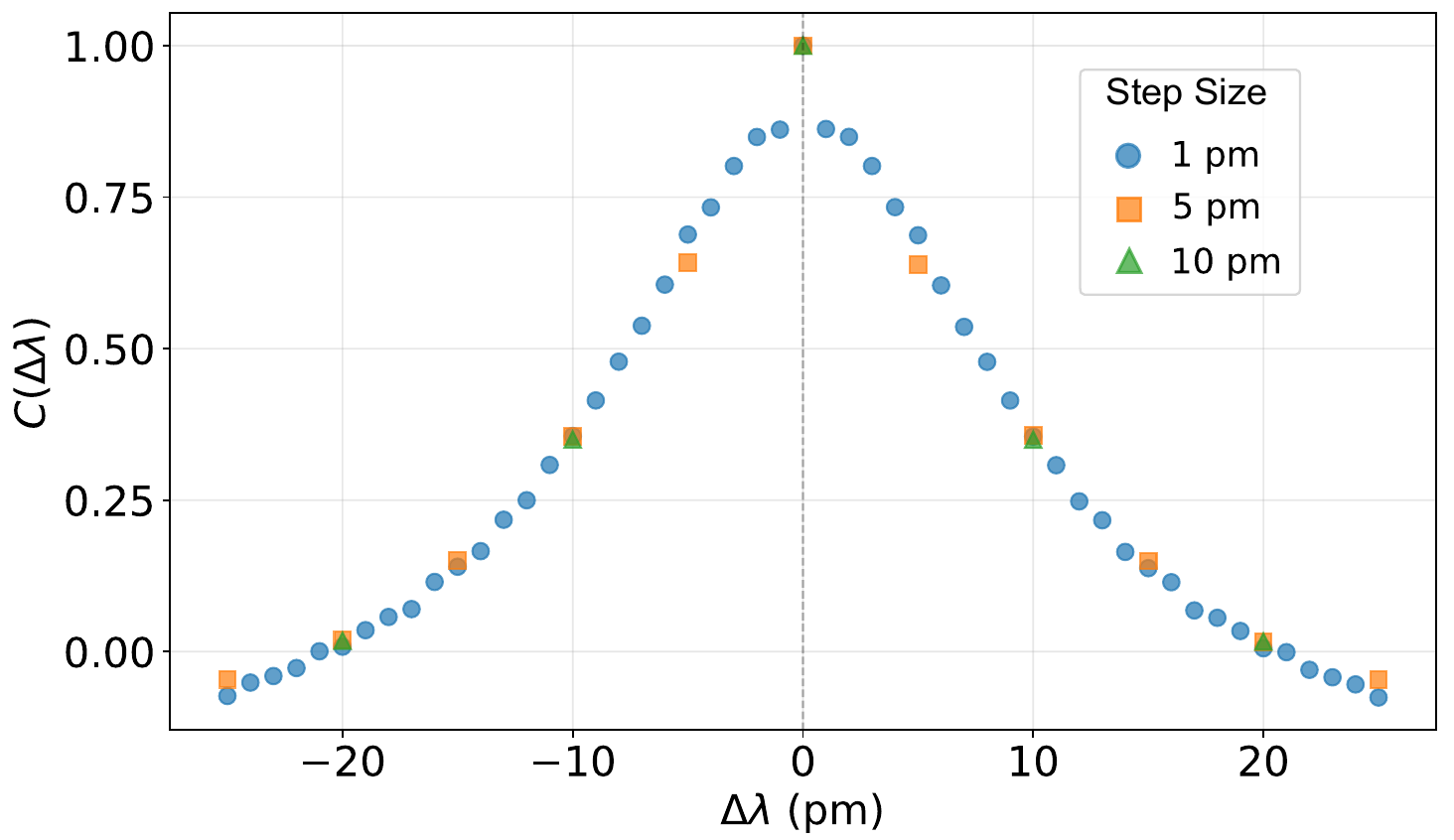}
    \caption{Spectral correlation curves for 1pm, 5pm and 10pm stepped wavelength scans.}
    \label{fig:optimum_spec_corr}
\end{figure}

\subsection*{S3: Stadium microresonator footprint}

Here, we investigate how scaling the surface area of the stadium microresonator affects the spectral correlation. 
We studied microcavities with surface areas ranging more than one order of magnitude, and all of them were etched for 6 minutes. 
For each device, the speckle's wavelength dependence was calculated scanning the C-band tunable laser sourrce with a 10pm step size. 
The spectral correlations were calculated and $\delta\lambda$ was extracted from the resulting curves. Across all fabricated devices, the extracted $\delta\lambda$ had a variance below 1pm, with no clear systematic trend on the microresonator's area. 
Notably, we could scale the footprint size of our stadium microcavities down to $0.01 \mathrm{pm}^2$ while keeping a spectral resolution of 7.0pm.

\begin{figure}[H]
    \centering
    \includegraphics[width=0.6\linewidth]{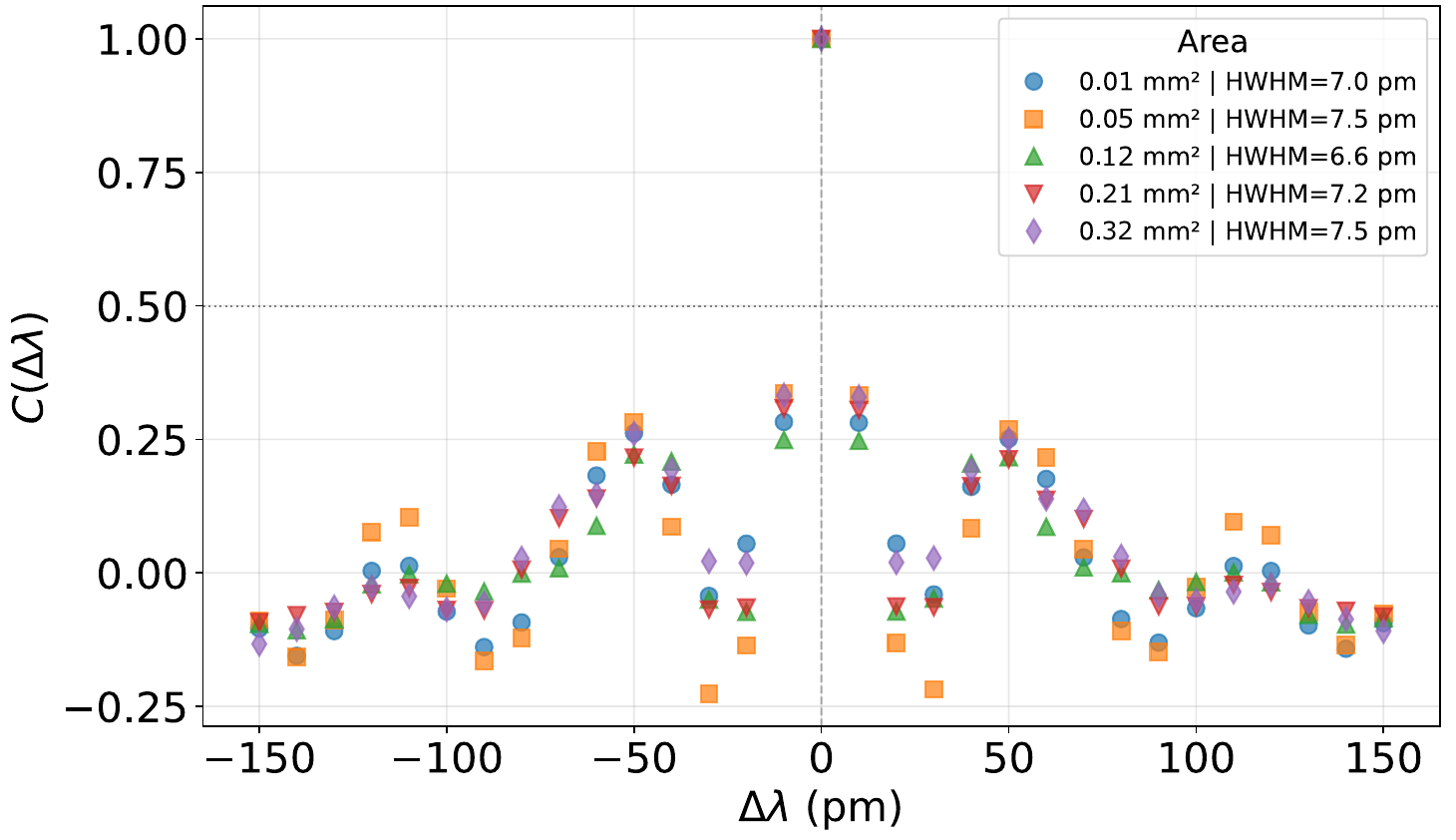}
    \caption{Spectral correlation decay curves for stadium microresonators with different surface areas.}
    \label{fig:area_scaling}
\end{figure}

\newpage

\subsection*{S4: Thermal Stability}

\begin{figure}[H]
    \centering
    \includegraphics[width=0.6\linewidth]{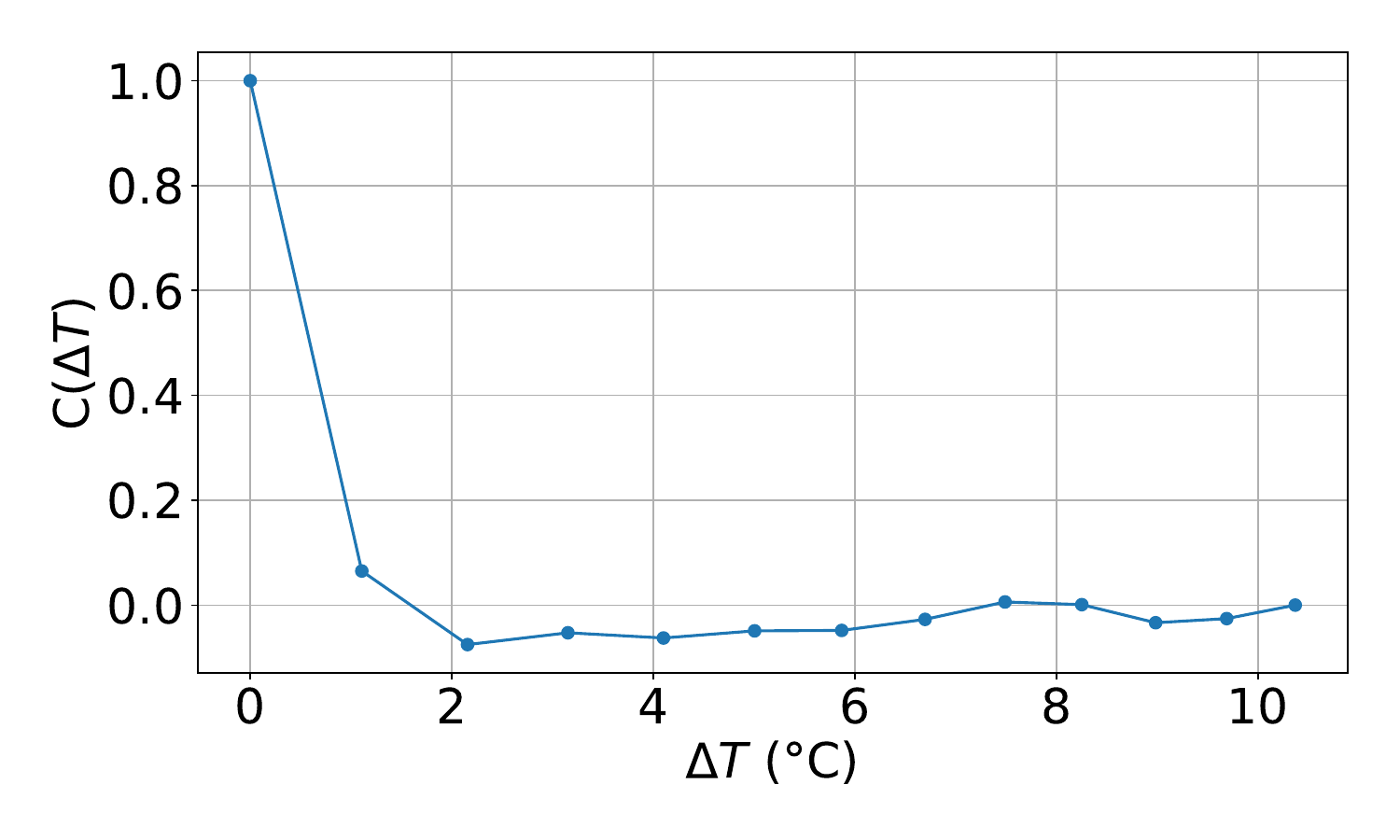}
    \caption{Speckle correlation as a function of temperature. 
    The stadium microcavity was injected with 1550nm light and the speckle pattern measured at different temperatures.}
    \label{fig:therm}
\end{figure}

We studied the thermal stability of the speckle patterns when injected at 1550nm. 
In this study, the photonic chip was placed on a stage with a Peltier cell and a 10k thermistor, connected to a temperature controller (Thorlabs, TED200C). 
The temperature controller controls the current flowing through the Peltier stage, which in turn provides a global temperature shift across the chip. 
We measured how the cross correlation changes from an image taken at a reference temperature shift of $\Delta T = 0^oC$ as we shift the temperature of the stage. 
Figure \ref{fig:therm} depicts the speckles correlation dependence with temperature $\mathrm{C}(\Delta T)$. 
We measured a half-width half-maximum drop of the correlation at $\sim 0.5^oC$. 

\subsection*{S5: Temporal stability of the speckle correlations}

\begin{figure}[H]
    \centering
    \includegraphics[width=0.6\linewidth]{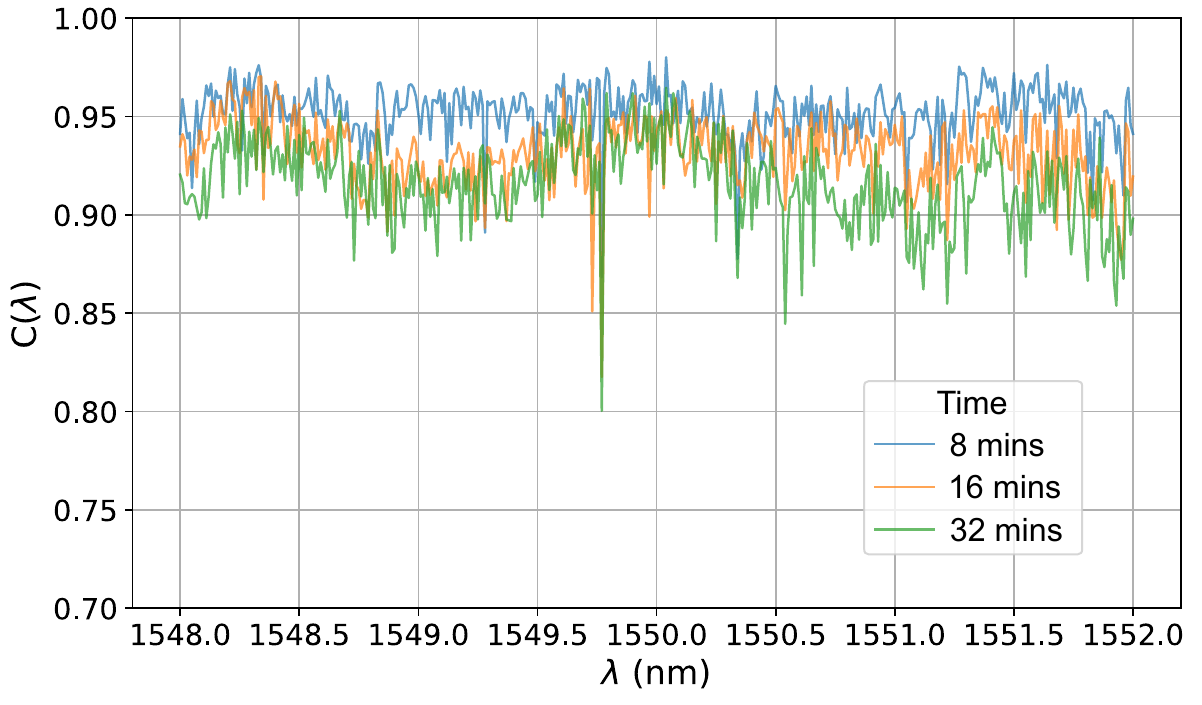}
    \caption{Speckle correlation between images taken at the same wavelengths during back to back sweeps for the IR tunable laser source. 
    No correlation drops below 80\% were observed over 32 minutes measurement (Each sweep took 5 minutes). 
    Gradual relaxation in the correlation between subsequent sweeps shows a time dependence on the stability of the measurement.}
    \label{fig:corr_stability_IR}
\end{figure}

Finally, we measured the repeatability of the speckle patterns over time. 
We performed cross-correlation measurements between equivalent speckle patterns acquired at the same wavelength during different sweeps for IR and for red. 
The IR stepped measurements took approx. 8 minutes per sweep, to measure a 4nm spectral window at a resolution of 0.01nm. 
The red swept measurements took approx. 100s per sweep to measure a 2nm spectral window at a resolution of 0.01nm. 
Figure \ref{fig:corr_stability_IR} corresponds to the IR measurement. 
We can observe that the speckles correlation oscillates within a 5\% value for all the range, with the average remaining above 90\% over the 32 minute measurement. 

For red wavelength we did not observe a correlation drop below 90\% over the 300s measurement window. 
In both cases, IR and red, we saw some decay in correlation over time which we attribute to environmental thermal fluctuations in the laboratory temperature. 
Those can be addressed packaging the spectrometer system (chip, optics and camera).

\begin{figure}[H]
    \centering
    \includegraphics[width=0.6\linewidth]{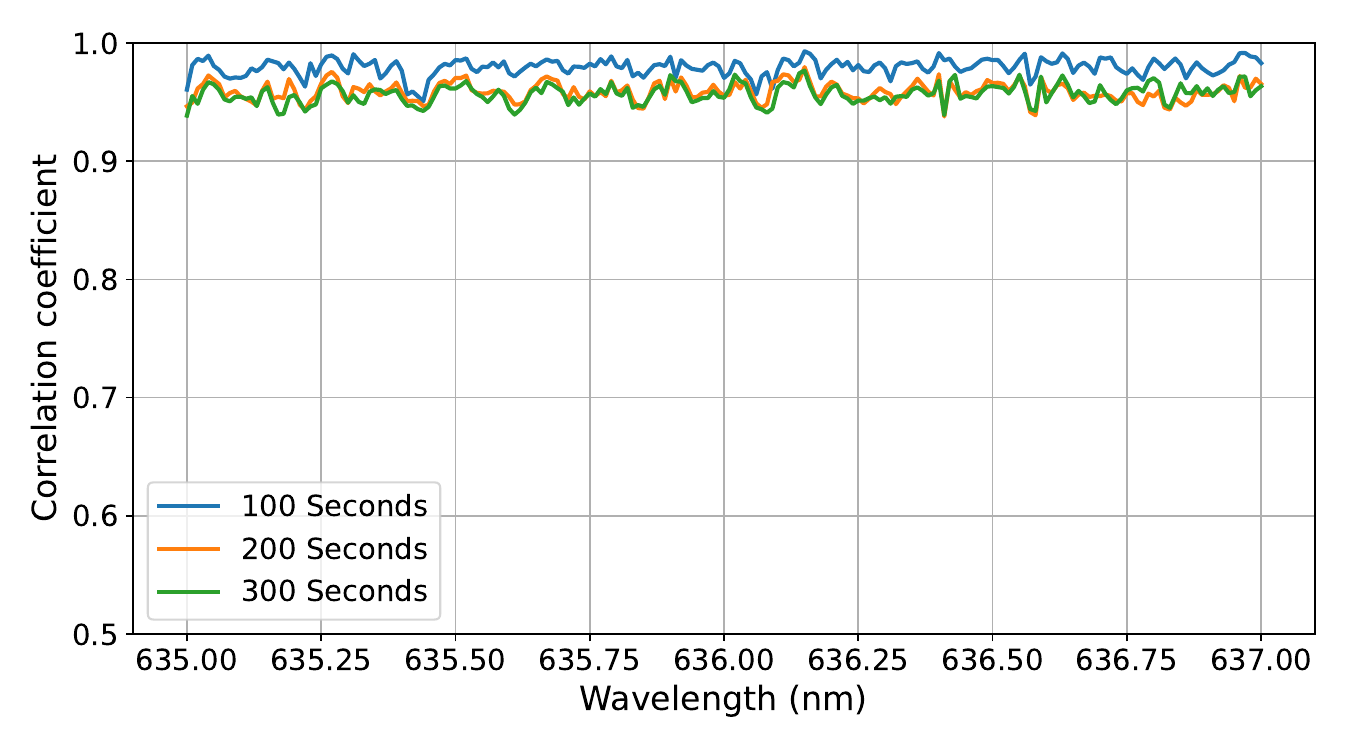}
    \caption{Speckle correlation between images taken at the same wavelength during back to back sweeps for the red tunable laser source. 
    Each sweep takes 100 seconds. 
    No correlation drops below 90\% although we acknowledge this is over a relatively short time-span. 
    Gradual relaxation in the correlation between subsequent sweeps shows a time dependence on the stability of the measurement. }
    \label{fig:corr_stability_red}
\end{figure}

\end{document}